\documentclass[aps,psfig,epsfig,preprint] {revtex4}
\usepackage{epsfig}
\usepackage{color}
\usepackage{bm}
\usepackage{subfigure}

\begin{document}
\draft
\title{
{\normalsize \hskip4.2in USTC-ICTS-07-08} \\{\bf Unparticle Physics
in DIS}}

\author{Gui-Jun Ding\footnote{E-mail: dinggj@mail.ustc.edu.cn}, Mu-Lin
Yan\footnote{E-mail: mlyan@ustc.edu.cn}}

\affiliation{\centerline{Interdisciplinary Center for Theoretical
Study and Department of Modern Physics,} \centerline{University of
Science and Technology of China, Hefei, Anhui 230026, China} }

\begin{abstract}
The unparticle stuff scenario related to the notrivial IR fixed
point in 4D-conformal field theory is recently suggested by Georgi.
We illustrate its physical effects in Deep Inelastic Scattering
(DIS) process. A possible signal of unparticle related to parity
violation asymmetry in DIS is investigated. It is found out that the
behavior of this parity violation signal is sensitive to the value
of  the scale dimension $d_{\cal U}$ of unpaticle.

\vskip 0.5cm

\end{abstract}
\maketitle
\section{introduction}
It is well-known that conformal symmetry plays important role in
both critical phenomena and the superstring theory. Nevertheless, in
particle physics in four dimensional space-time dimension, conformal
symmetry is broken by the masses of the particles explicitly, and
even though there is conformal symmetry classically, this symmetry
would be broken by the renormalization effect. However, at high
energy scale, there could exist stuff with notrivial scale
invariance in the infrared(such as the Banks-Zaks field\cite{bank}),
which is recently suggested by Howard Georgi as a component of the
beyond standard model(SM) physics above the TeV scale\cite{georgi1}.

Since unparticle stuff is completely new to us, and we don't have a
picture of what the unparticle looks like, and the theories with
notrivial IR fixed point is very complex, Georgi have used the
method of the low energy effective field theory to study the
unparticle stuff production\cite{georgi1} and the peculiar virtual
effects in high energy processes\cite{georgi2}. Moreover, the the
unparticle production in $e^{+}e^{-}\rightarrow\gamma{\cal U}$,
mono-jet production, the virtual effects of the unparticle in the
Drell-Yan process and the muon anomaly have been
investigated\cite{yuan}, and the fermionic unparticles are
introduced as well\cite{luo}. Unparticle physics have very
interesting and rich phenomenological consequences. In this paper,
we shall explore the phenomenological consequences of unparticle in
deep inelastic scattering(DIS) in the framework of the effective
field theory.

We follow closely the scenario studied in\cite{georgi1,georgi2}. The
fields of the theory with notrivial IR fixed point are denoted as
${\cal {BZ}}$ fields, and the SM fields and the ${\cal {BZ}}$ fields
interact through the exchanges of particles with a high mass scale
$M_{\cal U}$. Thus, below the high scale $M_{\cal U}$, the effective
norenormalizable couplings between them are follows,
\begin{equation}
\label{1}\frac{1}{M^{k}_{\cal U}}{\cal O}_{SM}{\cal O}_{\cal{BZ}},
\end{equation}
where ${\cal O}_{SM}$ and ${\cal O}_{\cal{BZ}}$ are respectively the
local operators built up of SM fields and the ${\cal {BZ}}$ fields.
Furthermore, below the scale $\Lambda_{\cal U}$ where the scale
invariance in the ${\cal {BZ}}$ sector emerges, the ${\cal {BZ}}$
operator ${\cal O}_{\cal{BZ}}$ is matched onto the unparticle
operator ${\cal O}_{\cal{U}}$, and Eq.(\ref{1}) is matched onto the
effective interactions between the SM fields and the unparticle
fields of the following form,
\begin{equation}
\label{2}\frac{C_{\cal U}\Lambda^{d_{\cal {BZ}}-d_{\cal U}}_{\cal
U}}{M^k_{\cal U}}{\cal O}_{SM}{\cal O}_{\cal{U}}
\end{equation}
where $d_{\cal {BZ}}$ and $d_{\cal U}$ are respectively the scale
dimensions of the local operators ${\cal O}_{\cal{BZ}}$ and ${\cal
O}_{\cal{U}}$, and $C_{\cal U}$ is the coefficient function which is
determined by the matching processes.

The following effective interactions of interesting phenomenologies
have been introduced in \cite{georgi1,georgi2},
\begin{eqnarray}
\nonumber&&\frac{C'_{S{\cal U}}\Lambda^{k-d_{\cal U}}_{\cal
U}}{M^k_{\cal U}}\;G_{\mu\nu}G^{\mu\nu}{\cal
O}_{\cal{U}}\;,~~~\frac{C_{T{\cal U}}\Lambda^{k-d_{\cal U}}_{\cal
U}}{M^k_{\cal U}}\;G_{\mu\lambda}G_{\nu}^{\lambda}{\cal
O}^{\mu\nu}_{\cal{U}}\;,~~~\frac{\Lambda^{k+1-d_{\cal U}}_{\cal
U}}{M^k_{\cal U}}\overline{f}\gamma_{\mu}(C_{V{\cal U}}+C_{A{\cal
U}}\gamma_5)f\;{\cal O}^{\mu}_{\cal{U}}\\
\label{3}&&\frac{\Lambda^{k+1-d_{\cal U}}_{\cal U}}{M^k_{\cal
U}}\overline{f}(C_{S{\cal U}}+iC_{P{\cal U}}\gamma_5)f\;{\cal
O}_{\cal{U}}
\end{eqnarray}
where $G_{\mu\nu}$ is the gluon field strength, $f$ denotes the SM
fermion fields, and  universal coupling between the unparticles and
the SM fermionic fields has been assumed. ${\cal O}_{\cal{U}}$,
${\cal O}^{\mu}_{\cal{U}}$ and ${\cal O}^{\mu\nu}_{\cal{U}}$ are
respectively the scalar, vector and tensor unparticle operators.
They are taken to be hermitian and the latter two operators are
assumed to be transverse. Naively $C_{V{\cal U}}$, $C_{A{\cal U}}$,
$C_{S{\cal U}}$ and $C_{P{\cal U}}$ should be of the same order. The
interactions in Eq.(\ref{3}) are generally weak, moreover the first
two couplings in Eq.(\ref{3}) are suppressed by
$\frac{1}{\Lambda_{\cal U}}$ in comparison with the latter two, so
the latter two effective interactions is dominant in the DIS
processes with virtual unparticle exchange. And we should work to
the lowest order in the small couplings of unpartile fields with the
SM fields in the effective theory. Similar to Ref.\cite{georgi2} the
following dimensionless coefficients are introduced for convenience,
\begin{equation}
\label{4}c_{V{\cal U}}=\frac{C_{V{\cal U}}\Lambda^{k+1-d_{\cal
U}}_{\cal U}}{M^k_{\cal U}M_Z^{1-d_{\cal U}}}\;,~~~c_{A{\cal
U}}=\frac{C_{A{\cal U}}\Lambda^{k+1-d_{\cal U}}_{\cal U}}{M^k_{\cal
U}M_Z^{1-d_{\cal U}}}\;~~~c_{S{\cal U}}=\frac{C_{S{\cal
U}}\Lambda^{k+1-d_{\cal U}}_{\cal U}}{M^k_{\cal U}M_Z^{1-d_{\cal
U}}}\;,~~~c_{P{\cal U}}=\frac{C_{P{\cal U}}\Lambda^{k+1-d_{\cal
U}}_{\cal U}}{M^k_{\cal U}M_Z^{1-d_{\cal U}}}
\end{equation}

Following Ref.\cite{georgi1,georgi2}, by conformal symmetry, we
have,
\begin{equation}
\label{5}\langle0|{\cal O}^{\mu}(0)|P\rangle\langle P|{\cal
O}^{\nu}(0)|0\rangle=A_{d_{\cal
U}}\;\theta(P^{0})\;\theta(P^2)(-g^{\mu\nu}+\frac{P^{\mu}P^{\nu}}{P^2})\;(P^2)^{d_{\cal
U}-2}
\end{equation}
where
\begin{equation}
\label{6}A_{d_{\cal U}}=\frac{16\pi^{5/2}}{(2\pi)^{2d_{\cal
U}}}\frac{\Gamma(d_{\cal U}+1/2)}{\Gamma(d_{\cal
U}-1)\Gamma(2d_{\cal U})}
\end{equation}
which is normalized with respect to the phase space of $d_{\cal U}$
massless particles, and the two-point correlation function of the
unparticle operator ${\cal O}^{\mu}$ can be obtained as follows,
\begin{equation}
\label{7}\int d^4x\;e^{iP\cdot x}\langle0|T({\cal O}^{\mu}(x)\;{\cal
O}^{\nu}(0))|0\rangle=\frac{iA_{d_{\cal U}}}{2\sin(d_{\cal
U}\pi)}\frac{-g^{\mu\nu}+P^{\mu}P^{\nu}/P^2}{(-P^2-i\epsilon)^{2-d_{\cal
U}}}
\end{equation}
Similarly,
\begin{equation}
\label{add1}\int d^4x\;e^{iP\cdot x}\langle0|T({\cal O}(x)\;{\cal
O}(0))|0\rangle=\frac{iA_{d_{\cal U}}}{2\sin(d_{\cal
U}\pi)}\frac{1}{(-P^2-i\epsilon)^{2-d_{\cal U}}}
\end{equation}

The above propagator factor has been obtained independently by
Georgi\cite{georgi2} and  Cheung {\it et al.,}\cite{yuan}.

\section{DIS processes and unparticle}
Since the effective interaction between the unparticle fields and
the SM fermions in Eq.(\ref{3}) is flavor-conserving, which is
consistent with the suppressed flavor changing neutral current
(FCNC) transitions, the unparticle will only affect the neutral
current ($\gamma$ and $Z$) exchange DIS processes
$\ell(\nu)N\rightarrow\ell(\nu)X$, which is shown in Fig.1. For the
charged lepton scattering $\ell N\rightarrow\ell X$, the
differential scattering cross section is,
\begin{equation}
\label{8}\frac{d^2\sigma^{\ell}}{dxdQ^2}=\frac{4\pi\alpha^2}{xQ^4}[\;xy^2F^{\ell}_1+(1-y)F^{\ell}_2+y(1-\frac{1}{2}y)xF^{\ell}_3\;]
\end{equation}

\begin{figure}[hptb]
\begin{center}
\includegraphics*[width=8cm]{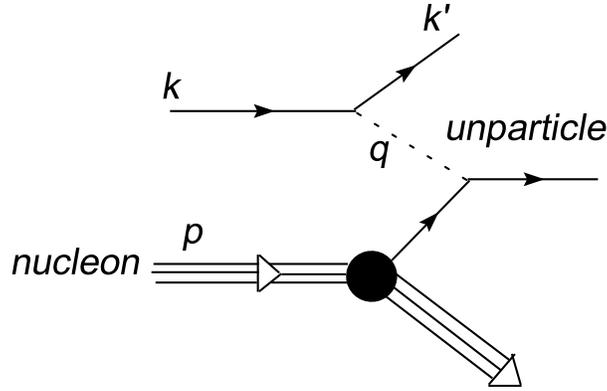}
\caption{Deep inelastic scattering with virtual unparticle exchange}
\end{center}
\label{fig1}
\end{figure}
\noindent where $F^{\ell}_i(x,Q^2)$(i=1,2,3) is the structure
functions which measure the structure of the target, and
\begin{eqnarray}
\nonumber
F^{\ell}_1(x,Q^2)&=&\sum_{q}[q(x)+\overline{q}(x)]\;B^{\ell}_q(Q^2)\\
\nonumber
F^{\ell}_2(x,Q^2)&=&\sum_{q}x\;[q(x)+\overline{q}(x)]\;C^{\ell}_q(Q^2)\\
\label{9}F^{\ell}_3(x,Q^2)&=&\sum_{q}[q(x)-\overline{q}(x)]\;D^{\ell}_q(Q^2)
\end{eqnarray}
with
\begin{eqnarray}
\nonumber
B^{\ell}_q(Q^2)&=&\frac{Q^2_q}{2}+\frac{(V^2_q+A^2_q)(V^2_{\ell}+A^2_{\ell})}{32\sin^4\theta_W\cos^4\theta_W}P^2_{Z}+\frac{(c^2_{V{\cal
U}}+c^2_{A{\cal U}})^2}{32\pi^2\alpha^2M^{4(d_{\cal
U}-1)}_Z}P^2_{{\cal
U}}-\frac{Q_qV_{q}V_{\ell}}{4\sin^2\theta_W\cos^2\theta_W}P_{Z}\\
\nonumber&&-\frac{Q_q\;c^2_{V{\cal U}}}{4\pi\alpha M^{2(d_{\cal
U}-1)}_Z}P_{{\cal U}}+\frac{(V_q\;c_{V{\cal U}}-A_q\;c_{A{\cal
U}})(V_{\ell}\;c_{V{\cal U}}-A_{\ell}\;c_{A{\cal U}})}{16\pi\alpha
\sin^2\theta_W\cos^2\theta_WM^{2(d_{\cal U}-1)}_Z}P_{{\cal
U}Z}\\
\label{add2}&&+\frac{(c^2_{S{\cal U}}-c^2_{P{\cal
U}})^2}{64\pi^2\alpha^2M^{4(d_{\cal U}-1)}_Z}P^2_{{\cal
U}}\\
\nonumber
C^{\ell}_q(Q^2)&=&Q^2_q+\frac{(V^2_q+A^2_q)(V^2_{\ell}+A^2_{\ell})}{16\sin^4\theta_W\cos^4\theta_W}P^2_{Z}+\frac{(c^2_{V{\cal
U}}+c^2_{A{\cal U}})^2}{16\pi^2\alpha^2M^{4(d_{\cal
U}-1)}_Z}P^2_{{\cal
U}}-\frac{Q_qV_{q}V_{\ell}}{2\sin^2\theta_W\cos^2\theta_W}P_{Z}\\
\label{10}&&-\frac{Q_q\;c^2_{V{\cal U}}}{2\pi\alpha M^{2(d_{\cal
U}-1)}_Z}P_{{\cal U}}+\frac{(V_q\;c_{V{\cal U}}-A_q\;c_{A{\cal
U}})(V_{\ell}\;c_{V{\cal U}}-A_{\ell}\;c_{A{\cal U}})}{8\pi\alpha
\sin^2\theta_W\cos^2\theta_WM^{2(d_{\cal U}-1)}_Z}P_{{\cal U}Z}\\
\nonumber
D^{\ell}_q(Q^2)&=&\frac{V_qV_{\ell}A_{q}A_{\ell}}{4\sin^4\theta_W\cos^4\theta_W}P^2_{Z}+\frac{c^2_{V{\cal
U}}\;c^2_{A{\cal U}}}{4\pi^2\alpha^2M^{4(d_{\cal U}-1)}_Z}P^2_{\cal
U}-\frac{Q_{q}A_{q}A_{\ell}}{2\sin^2\theta_W\cos^2\theta_W}P_{Z}\\
\label{11}&&-\frac{Q_q\;c^2_{A{\cal U}}}{2\pi\alpha M^{2(d_{\cal
U}-1)}_Z}P_{\cal U}+\frac{(V_q\;c_{A{\cal U}}-A_q\;c_{V{\cal
U}})(V_{\ell}\;c_{A{\cal U}}-A_{\ell}\;c_{V{\cal U}})}{8\pi\alpha
\sin^2\theta_W\cos^2\theta_WM^{2(d_{\cal U}-1)}_Z}P_{{\cal U}Z}\\
\label{12} P_{Z}&=&\frac{Q^2}{Q^2+M^2_Z}\;,~~~P_{\cal
U}=\frac{A_{d_{\cal U}}}{2\sin(d_{\cal U}\pi)}(Q^2)^{(d_{\cal
U}-1)}\;,~~~P_{{\cal U}Z}=\frac{A_{d_{\cal U}}}{2\sin(d_{\cal
U}\pi)}\frac{(Q^2)^{d_{\cal U}}}{Q^2+M^2_Z}
\end{eqnarray}
In Eq.(\ref{add2}-\ref{11}), $\theta_{W}$ is the Weinberg angle
$\sin^2\theta_W\simeq0.23$\cite{pdg}, $M_Z$ is the mass of the
$Z^{0}$ gauge boson $M_Z\simeq91.19$GeV\cite{pdg}, and
$V_f=T_{3f}-2Q_{f}\sin^2\theta_W$, $A_{f}=T_{3f}$, where $Q_f$ is
the electric charge of the fermion $f$ in unit of the positron
electric charge $e$. $Q^2$, $x$ and $y$ are the standard deep
inelastic variables, which are defined by:
\begin{equation}
\label{13}Q^{2}=-(k-k')^2=2k\cdot
k'\;,~~x=\frac{Q^2}{2p\cdot(k-k')}\;,~~y=\frac{p\cdot(k-k')}{p\cdot
k}
\end{equation}

The $Callan-Gross$ relation $F^{\ell}_2(x,Q^2)=2xF^{\ell}_1(x,Q^2)$
is violated by the term $\frac{(c^2_{S{\cal U}}-c^2_{P{\cal
U}})^2}{64\pi^2\alpha^2M^{4(d_{\cal U}-1)}_Z}P^2_{{\cal U}}$ in
Eq.(\ref{add2}) at the present stage, and it will receive additional
small corrections if we include the suppressed effective
interactions between the unparticle and the gluon in Eq.(\ref{3}).
If we set $c_{V{\cal U}}=c_{A{\cal U}}=c_{S{\cal U}}=c_{P{\cal
U}}=0$, ( ${\it i.e.,}$ if the unparticle stuff doesn't exist in the
Nature) the differential cross section Eq.(\ref{8}) coincides with
the well-known results. When $c_{i{\cal U}}\neq 0$(i=V, A, S, P),
the contributions to the structure functions
$F^{\ell}_i(x,Q^2)$(i=1,2,3) due to unparticle emerge. Since the
coupling $c_{V{\cal U}}$, $c_{A{\cal U}}$, $c_{S{\cal U}}$ and
$c_{P{\cal U}}$ are small, to the lowest nontrivial order, the
corrections to the structure functions are the interference terms
between the space-like vector unparticle exchange amplitudes and the
standard model amplitudes, whereas the leading corrections to
$e^{+}e^{-}\rightarrow\mu^{+}\mu^{-}$ are the interference terms
between the time-like vector unparticle exchange amplitudes and the
standard model amplitudes\cite{georgi2}.

For the neutrino scattering $\nu N\rightarrow\nu X$, the
corresponding differential scattering cross section is,
\begin{equation}
\label{14}\frac{d^2\sigma^{\nu}}{dxdQ^2}=\frac{G^2_F}{2\pi
x}(\frac{M^2_Z}{Q^2+M^2_Z})^2[\;xy^2F^{\nu}_1+(1-y)F^{\nu}_2+y(1-\frac{1}{2}y)xF^{\nu}_3\;]
\end{equation}
the structure functions $F^{\nu}_i(x,Q^2)$(i=1,2,3) are as follows,
\begin{eqnarray}
\nonumber F_1(x,Q^2)&=&\sum_{q}\;[q(x)+\overline{q}(x)]B^{\nu}_q(Q^2)\\
\nonumber
F^{\nu}_2(x,Q^2)&=&\sum_{q}x[q(x)+\overline{q}(x)]C^{\nu}_q(Q^2)\\
\label{15}
F^{\nu}_3(x,Q^2)&=&\sum_{q}[q(x)-\overline{q}(x)]D^{\nu}_q(Q^2)
\end{eqnarray}
with
\begin{eqnarray}
\nonumber B^{\nu}_q(Q^2)&=&\frac{V^2_q+A^2_q}{2}+\frac{(c^2_{V{\cal
U}}+c^2_{A{\cal U}})^2}{2M^{4d_{\cal U}}_ZG^2_F}R^2_{{\cal
U}Z}+\frac{(V_q\;c_{V{\cal U}}-A_q\;c_{A{\cal U}})(c_{V{\cal
U}}-c_{A{\cal U}})}{\sqrt{2}\;M^{2d_{\cal U}}_ZG_F}R_{{\cal U}Z}\\
\label{add3}&&+\frac{(c^2_{S{\cal U}}-c^2_{P{\cal
U}})^2}{4M^{4d_{\cal U}}_ZG^2_F}R^2_{{\cal U}Z}\\
\label{16}C^{\nu}_q(Q^2)&=&V^2_q+A^2_q+\frac{(c^2_{V{\cal
U}}+c^2_{A{\cal U}})^2}{M^{4d_{\cal U}}_ZG^2_F}R^2_{{\cal
U}Z}+\frac{\sqrt{2}\;(V_q\;c_{V{\cal U}}-A_q\;c_{A{\cal
U}})(c_{V{\cal
U}}-c_{A{\cal U}})}{M^{2d_{\cal U}}_ZG_F}R_{{\cal U}Z}\\
\label{17}D^{\nu}_q(Q^2)&=&2V_qA_{q}+\frac{4c^2_{V{\cal
U}}\;c^2_{A{\cal U}}}{M^{4d_{\cal U}}_ZG^2_F}R^2_{{\cal
U}Z}-\frac{\sqrt{2}\;(V_q\;c_{A{\cal U}}-A_q\;c_{V{\cal
U}})(c_{V{\cal
U}}-c_{A{\cal U}})}{M^{2d_{\cal U}}_ZG_F}R_{{\cal U}Z}\\
\label{18}R_{{\cal U}Z}&=&\frac{A_{d_{\cal U}}}{2\sin(d_{\cal
U}\pi)}(Q^2)^{d_{\cal U}-2}(Q^2+M^2_Z)
\end{eqnarray}
where $G_F$ is the Fermi constant
$G_F\simeq1.166\times10^{-5}\rm{GeV}^2$\cite{pdg}. If we set
$c_{V{\cal U}}=c_{A{\cal U}}=c_{S{\cal U}}=c_{P{\cal U}}=0$, the
differential cross section reduces to the well-known result, and the
$Callan-Gross$ relation $F^{\nu}_2(x,Q^2)=2xF^{\nu}_1(x,Q^2)$ is
violated by the $\frac{(c^2_{S{\cal U}}-c^2_{P{\cal
U}})^2}{4M^{4d_{\cal U}}_ZG^2_F}R^2_{{\cal U}Z}$ term in
Eq.(\ref{add3}).

In order to see how the unparticle affects the structure functions,
it is instructive to first assume $c_{V{\cal U}}=0$, then there
isn't interference between the unparticle exchange amplitudes and
the photon exchange amplitudes. We can easily see that dominant
correction is proportional to $c^2_{A{\cal U}}$, here we omit the
high order terms which contain $c^4_{A{\cal U}}$, $c^4_{S{\cal U}}$
and $c^4_{P{\cal U}}$, and in this case the changes of the functions
$B^{\ell}_q(Q^2)$, $C^{\ell}_q(Q^2)$, $D^{\ell}_q(Q^2)$,
$B^{\nu}_q(Q^2)$ $C^{\nu}_q(Q^2)$ and $D^{\nu}_q(Q^2)$ caused by
unparticle exchanges are respectively the following:
\begin{eqnarray}
\nonumber 2\Delta B^{\ell}_q(Q^2)/c^2_{A{\cal U}}&\approx&\Delta
C^{\ell}_q(Q^2)/c^2_{A{\cal U}}\approx\frac{A_qA_{\ell}}{16\pi\alpha
M^{2(d_{\cal U}-1)}_Z\sin^2\theta_W\cos^2\theta_W}\frac{A_{d_{\cal
U}}}{\sin(d_{\cal U}\pi)}\frac{(Q^2)^{d_{\cal U}}}{Q^2+M^2_Z}\\
\nonumber\Delta D^{\ell}_q(Q^2)/c^2_{A{\cal
U}}&\approx&-\frac{Q_q}{4\pi\alpha M^{2(d_{\cal
U}-1)}_Z}\frac{A_{d_{\cal U}}}{\sin(d_{\cal U}\pi)}(Q^2)^{(d_{\cal
U}-1)}+\frac{V_qV_{\ell}}{16\pi\alpha\sin^2\theta_W\cos^2\theta_WM^{2(d_{\cal
U}-1)}_Z}\\
\nonumber&&\times\frac{A_{d_{\cal U}}}{\sin(d_{\cal
U}\pi)}\frac{(Q^2)^{d_{\cal U}}}{Q^2+M^2_Z}\\
\nonumber 2\Delta B^{\nu}_q(Q^2)/c^2_{A{\cal U}}&\approx&\Delta
C^{\nu}_q(Q^2)/c^2_{A{\cal U}}\approx\frac{A_q}{\sqrt{2}M^{2d_{\cal
U}}_Z G_F}\frac{A_{d_{\cal
U}}}{\sin(d_{\cal U}\pi)}(Q^2)^{(d_{\cal U}-2)}(Q^2+M^2_Z)\\
\label{19}\Delta D^{\nu}_q(Q^2)/c^2_{A{\cal
U}}&\approx&\frac{V_q}{A_q}(\Delta C^{\nu}_q(Q^2)/c^2_{A{\cal U}})
\end{eqnarray}

As a illustration the leading order unparticle corrections $\Delta
C^{e}_u(Q^2)/c^2_{A{\cal U}}$, $\Delta D^{e}_u(Q^2)/c^2_{A{\cal U}}$
and  $\Delta C^{\nu}_u(Q^2)/c^2_{A{\cal U}}$ for various $d_{\cal
U}$ are respectively shown in Fig.2-Fig.4. Since $\Delta
D^{\nu}_q(Q^2)/c^2_{A{\cal U}}$ is proportional to $\Delta
C^{\nu}_q(Q^2)/c^2_{A{\cal U}}$, we have not shown the profile of
$\Delta D^{\nu}_q(Q^2)/c^2_{A{\cal U}}$. For $1<d_{\cal U}<2$,
$\sin(d_{\cal U}\pi)$ is negative, so $\Delta
C^{\nu}_u(Q^2)/c^2_{A{\cal U}}$ shown in Fig.4 is negative, while
$\Delta C^{e}_u(Q^2)/c^2_{A{\cal U}}$ and $\Delta
D^{e}_u(Q^2)/c^2_{A{\cal U}}$ are positive. We note that $\Delta
C^{e}_u(Q^2)/c^2_{A{\cal U}}$ and $\Delta D^{e}_u(Q^2)/c^2_{A{\cal
U}}$ increase with the increase of $Q^2$, however, $|\Delta
C^{\nu}_u(Q^2)/c^2_{A{\cal U}}|$ firstly decrease, then begin to
increase at $Q^2=\frac{2-d_{\cal U}}{d_{\cal U}-1}M^2_Z$.

After discussion on the pure axial vector unparticle couplings, we
now turn to the pure vector coupling case, {\it i.e, }
 $c_{V{\cal U}}$ is nozero and $c_{A{\cal U}}=0$. The leading order
corrections $\Delta B^{\ell}_q(Q^2)/c^2_{V{\cal U}}$, $\Delta
C^{\ell}_q(Q^2)/c^2_{V{\cal U}}$, $\Delta
D^{\ell}_q(Q^2)/c^2_{V{\cal U}}$, $\Delta B^{\nu}_q(Q^2)/c^2_{V{\cal
U}}$, $\Delta C^{\nu}_q(Q^2)/c^2_{V{\cal U}}$ and $\Delta
D^{\nu}_q(Q^2)/c^2_{V{\cal U}}$ induced by the unparticle in this
case are as follows:
\begin{eqnarray}
\nonumber 2\Delta B^{\ell}_q(Q^2)/c^2_{V{\cal U}}&\approx&\Delta
C^{\ell}_q(Q^2)/c^2_{V{\cal U}}\approx-\frac{Q_q}{4\pi\alpha
M^{2(d_{\cal U}-1)}_Z}\frac{A_{d_{\cal U}}}{\sin(d_{\cal
U}\pi)}(Q^2)^{(d_{\cal U}-1)}\\
\nonumber&&+\frac{V_qV_{\ell}}{16\pi\alpha\sin^2\theta_W\cos^2\theta_W
M^{2(d_{\cal U}-1)}_Z} \frac{A_{d_{\cal U}}}{\sin(d_{\cal
U}\pi)}\frac{(Q^2)^{d_{\cal U}}}{Q^2+M^2_Z}\\
\nonumber\Delta D^{\ell}_q(Q^2)/c^2_{V{\cal
U}}&\approx&\frac{A_qA_{\ell}}{16\pi\alpha\sin^2\theta_W\cos^2\theta_W
M^{2(d_{\cal U}-1)}_Z}\frac{A_{d_{\cal
U}}}{\sin(d_{\cal U}\pi)}\frac{(Q^2)^{d_{\cal U}}}{Q^2+M^2_Z}\\
\nonumber 2\Delta B^{\nu}_q(Q^2)/c^2_{V{\cal U}}&\approx&\Delta
C^{\nu}_q(Q^2)/c^2_{V{\cal U}}\approx\frac{V_q}{\sqrt{2}M^{2d_{\cal
U}}_Z G_F}\frac{A_{d_{\cal
U}}}{\sin(d_{\cal U}\pi)}(Q^2)^{(d_{\cal U}-2)}(Q^2+M^2_Z)\\
\label{20}\Delta D^{\nu}_q(Q^2)/c^2_{V{\cal
U}}&\approx&\frac{A_q}{V_q}(\Delta C^{\nu}_q(Q^2)/c^2_{V{\cal U}})
\end{eqnarray}

From the above equations, we can see that $\Delta B^{\ell}_q(Q^2)$,
$\Delta C^{\ell}_q(Q^2)$, $\Delta D^{\ell}_q(Q^2)$,  $\Delta
B^{\nu}_q(Q^2)$, $\Delta C^{\nu}_q(Q^2)$ and $\Delta D^{\nu}_q(Q^2)$
in the pure vector unparticle coupling case are closely related to
those in the pure axial vector coupling case. Thus, we can learn how
$\Delta C^{e}_u(Q^2)/c^2_{V{\cal U}}$, $\Delta
D^{e}_u(Q^2)/c^2_{V{\cal U}}$ and  $\Delta
C^{\nu}_u(Q^2)/c^2_{V{\cal U}}$ vary  with respect to $Q^2$ in the
pure vector coupling case by making some replacements in Fig.2-Fig.4
.

\section{asymmetries in deep inelastic polarized electron nucleon scattering and unparticle}

There is a strong belief in the physics community that the Standard
Model of particles and interactions is incomplete. Much effort has
been paid to search for  new physics, and will continue to do so in
the upgraded Tevatron, the LHC, and the future Linear Collider(ILC).
Direct searches are complemented by precision electro-weak
experiments that search for the indirect effects of new physics by
comparison with expectations calculable in the Standard Model.
Parity violation as an important indirect search for new physics
provides precise measurement of electroweak couplings at low $Q^2$.
This measurement is complementary to other existing or planned
precision measurement, then yields strong constraints on the
possible deviations from the Standard Model predictions and
distinguish various new physics. The Jefferson laboratory has planed
to measure precisely the parity violating asymmetry in
DIS\cite{jlab}.

The general four fermion lagrangian takes the following
form\cite{musolf}, which is responsible for the new physics
contribution to the parity violation DIS asymmetry,
\begin{equation}
\label{21}{\cal L}_{PV}=\frac{4\pi
\kappa^2}{\Lambda^2}\sum_{q,i,j}h^{q}_{ij}\overline{e}_i\gamma_{\mu}e_i\overline{q}_j\gamma^{\mu}q_{j}
\end{equation}
where $\kappa$ is the coupling strength of the new interaction,
$\Lambda$ is the characteristic mass of the new degree of freedom,
and the $h^q_{ij}$ are the helicity-dependent coupling parameters of
the quark $q$ with $i$ and $j$ denoting the handedness of the given
fermion. From the effective interaction Eq.(\ref{3}) between
unparticle and the SM field, we learn that the virtual exchange of
unparticle can result in the following four fermion interaction
relevant to the parity violating DIS asymmetry.
\begin{eqnarray}
\nonumber{\cal L}^{\cal{U}}_{PV}&=&\frac{A_{d_{\cal
U}}}{2\sin{(d_{\cal
U}}\pi)}\frac{1}{Q^2}(\frac{Q^2}{M^2_Z})^{d_{\cal
U}-1}\sum_q[c^2_{R{\cal U}}\;\overline{e}\gamma_{\mu}P_R
e\overline{q}\gamma^{\mu}P_R{q}+c_{R{\cal U}}c_{L{\cal
U}}\;\overline{e}\gamma_{\mu}P_R e\overline{q}\gamma^{\mu}P_L{q}\\
\label{22}&&+c_{R{\cal U}}c_{L{\cal U}}\;\overline{e}\gamma_{\mu}P_L
e\overline{q}\gamma^{\mu}P_R{q}+c^2_{L{\cal
U}}\;\overline{e}\gamma_{\mu}P_L e\overline{q}\gamma^{\mu}P_L{q}]
\end{eqnarray}
here $P_R$ and $P_L$ are the usual projection operator
$P_{R,L}=\frac{1}{2}(1\pm\gamma_5)$, $c_{R{\cal U}}$ and $c_{L{\cal
U}}$ are expressed in terms of the vector and axial vector coupling
constants $c_{V{\cal U}}$, $c_{A{\cal U}}$,
\begin{eqnarray}
\nonumber c_{R{\cal U}}&=&c_{V{\cal U}}+c_{A{\cal U}}\\
\label{23}c_{L{\cal U}}&=&c_{V{\cal U}}-c_{A{\cal U}}
\end{eqnarray}
For a isosinglet target such as the deuteron, the assumption of
isospin symmetry is generally made (i.e., all u and d distributions
interchanged for the proton and neutron). At sufficiently high $x$,
the relative importance of sea quark contributions approaches zero.
And then the parity asymmetry $A_{ed}(x,y)$ is insensitive to parton
distribution functions. To the leading order of $Q^2/M^2_Z$, after
length calculation, we have
\begin{eqnarray}
\nonumber
A_{ed}(x,y)&=&\frac{\sigma_R-\sigma_L}{\sigma_R+\sigma_L}=-\frac{G_FQ^2}{2\sqrt{2}\pi\alpha}\frac{9}{10}\{[1-\frac{20}{9}\sin^2\theta_W+\frac{1}{3}\xi(Q^2)\;c_{V{\cal
U}}c_{A{\cal U}}]+[1-4\sin^2\theta_W+\\
\label{24}&&\frac{1}{3}\xi(Q^2)\;
c_{V{\cal U}}c_{A{\cal
U}}]\frac{1-(1-y)^2}{1+(1-y)^2}\}
\end{eqnarray}
where $\xi(Q^2)=\frac{\sqrt{2}A_{d_{\cal U}}}{\sin(d_{\cal
U}\pi)}\frac{1}{Q^2G_F}(\frac{Q^2}{M^2_Z})^{d_{\cal U}-1}$, if we
set $c_{V{\cal U}}=c_{A{\cal U}}=0$, the above asymmetry factor
$A_{ed}(x,y)$ reduces to the well-known results in SM\cite{musolf}.
As an illustration, the predicted asymmetries in deep inelastic
electron-deuteron scattering are shown in Fig.5 for
$Q^2=10\rm{GeV}^2$ and with different scale dimensions of
unparticle. From this figure, we can see that the asymmetry $A_{ed}$
is very sensitive to the scale dimension $d_{\cal U}$: if
$1.5<d_{\cal U}<2$  we almost can not distinguish the SM from the
new physics with unparticle, however when $1<d_{\cal U}<1.5$, the
situation is changed drastically, i.e., the differences between
$A_{ed}$ due to unparticle and one of SM become rather large.
Consequently, by this result, the future precision measurement of
$A_{ed}$ would impose a strong constrain to $d_{\cal U}$.

For the sake of completeness we now indicate the result for a proton
target, rather than give qualitative results for all x and y, we
have chosen $x=\frac{1}{3}$ to give quantitative predictions. Here
the "valence" quarks dominate, and furthermore for the proton
$u(x=\frac{1}{3})\approx d(x=\frac{1}{3})$,as in the most naive
quark-parton model. Then  we find the parity asymmetry
$A_{ep}(x=\frac{1}{3},y)$ in deep inelastic electron-proton
scattering,
\begin{eqnarray}
\nonumber
A_{ep}(x=\frac{1}{3},y)&=&\frac{\sigma_R-\sigma_L}{\sigma_R+\sigma_L}=-\frac{G_FQ^2}{2\sqrt{2}\pi\alpha}\{[\frac{5}{6}-2\sin^2\theta_W+\frac{1}{2}\xi(Q^2)\;c_{V{\cal
U}}c_{A{\cal U}}]+[\frac{5}{6}(1-4\sin^2\theta_W)\\
\label{24}&&+\frac{1}{2}\xi(Q^2)\; c_{V{\cal U}}c_{A{\cal
U}}]\frac{1-(1-y)^2}{1+(1-y)^2}\}
\end{eqnarray}

The above parity asymmetry $A_{ep}(x=\frac{1}{3},y)$ becomes the
standard result in SM in the case of $ c_{V{\cal U}}=c_{A{\cal
U}}=0$\cite{musolf}. The asymmetry predicted in SM and new physics
with unparticle is shown in Fig.6 for longitudinally polarized
electron-proton deep inelastic scattering. Similar to the
electron-deuteron case, it is always negative, and the asymmetry
$A_{ep}(x=\frac{1}{3},y)$ in SM almost coincide with that in the
unparticle scenario for $1.5<d_{\cal U}<2$, but not as large (in
magnitude) as for the electron-deuteron case.

In Fig. 5 and 6, in order to show the $A_{ep}(x=\frac{1}{3},y)$
curves explicitly and concretely, we have taken $c_{V{\cal
U}}=c_{A{\cal U}}=0.01$. The values of $c_{V{\cal U}}$ and
$c_{A{\cal U}}$ reflect the strength of coupling between
SM-operators and the unparticle's. Since the new physics corrections
to SM due to unparticle must be rather small, we have $|c_{V{\cal
U}}|\sim |c_{A{\cal U}}|<<1$. When $c_{V{\cal U}}=c_{A{\cal
U}}=0.01$ is assumed, the corrections to $A_{ep}(x=\frac{1}{3},y)$
are proportional to order of ${\mathcal O}(c_{V{\cal U}}c_{A{\cal
U}})\sim 10^{-4}.$

\section{conclusion}

Unparticle stuff with nontrivial scale invariance may exist in our
world, and there are a lot of rich phenomenologies associated with
the unparticle, such as the unparticle physics effects in
fragmentation functions and its contribution to the NuTeV anamoly
{\it etc}\cite{ding}, which can serve as the experimental tests of
the unparticle. In this paper, we have demonstrated the unparticle
effects in the neutral current exchange DIS processes
$\ell(\nu)N\rightarrow\ell(\nu)X$ and the parity violation asymmetry
in electron nucleon deep inelastic scattering in the framework of
the effective theory.

Low energy experiment is uniquely sensitive to new physics, which is
complementary to direct new physics searches and very useful to
distinguish various new physics pictures. If there is really a scale
invariant sector, it would manifest itself not only in parity
violation asymmetry in electron-nucleon deep inelastic scattering,
but also in parity violation M$\ddot{\rm{o}}$ller scattering, atomic
parity violation, parity violation electron-proton scattering {\it
etc} low energy measurements. Parity violating asymmetry in DIS is
investigated in this work, together with precise measurement of
other low energy observables (such as the QWeak experiment at JLab),
they will yield strong constraints on the unparticle parameter, and
will play important role in distinguish the unparticle scenario from
other extensions of the SM(such as supersymmetry, leptoquarks and so
on).

\section *{ACKNOWLEDGEMENTS}
\indent This work is partially supported by National Natural Science
Foundation of China under Grant Numbers 90403021,  and KJCX2-SW-N10
of the Chinese Academy.

\begin{figure}[hptb]
\begin{center}
\includegraphics*[width=8cm]{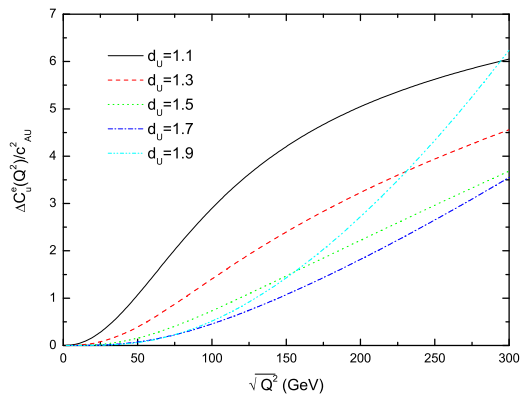}
\caption{ The unparticle correction to $C^{e}_u(Q^2)$ in unit of
$c^2_{A{\cal U}}$ for $d_{\cal U}=1.1, 1.3, 1.5, 1.7, 1.9$ in the
$c_{V{\cal U}}=0$ and $c_{A{\cal U}}\neq0$ case }
\end{center}
\label{fig2}
\end{figure}

\begin{figure}[hptb]
\begin{center}
\includegraphics*[width=8cm]{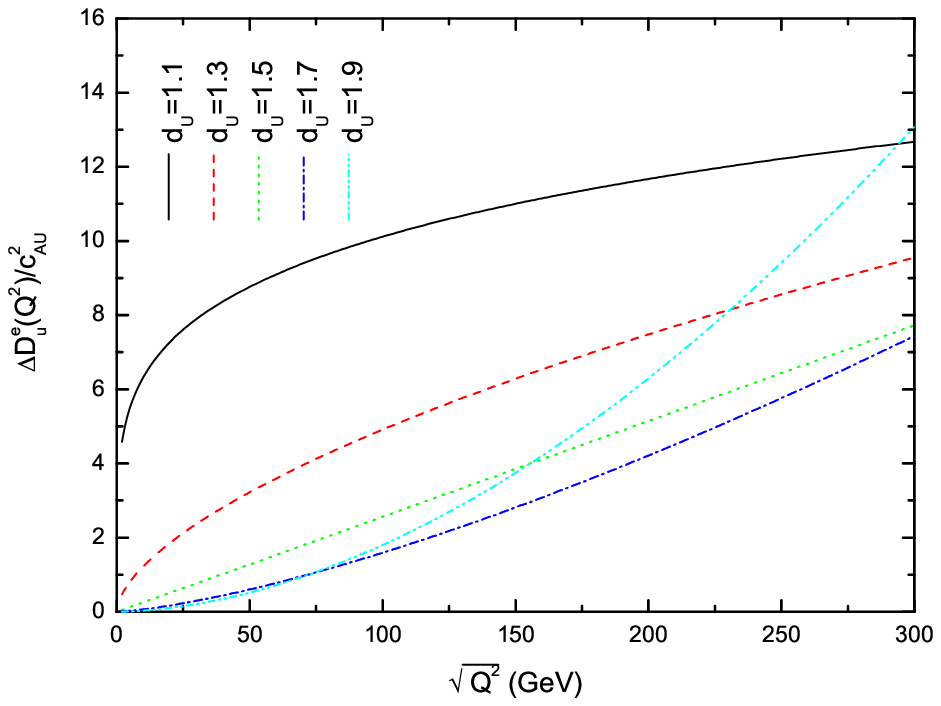}
\caption{The unparticle correction to $D^{e}_u(Q^2)$ in unit of
$c^2_{A{\cal U}}$ for $d_{\cal U}=1.1, 1.3, 1.5, 1.7, 1.9$ in the
$c_{V{\cal U}}=0$ and $c_{A{\cal U}}\neq0$ case}
\end{center}
\label{fig3}
\end{figure}

\begin{figure}[hptb]
\begin{center}
\includegraphics*[width=8cm]{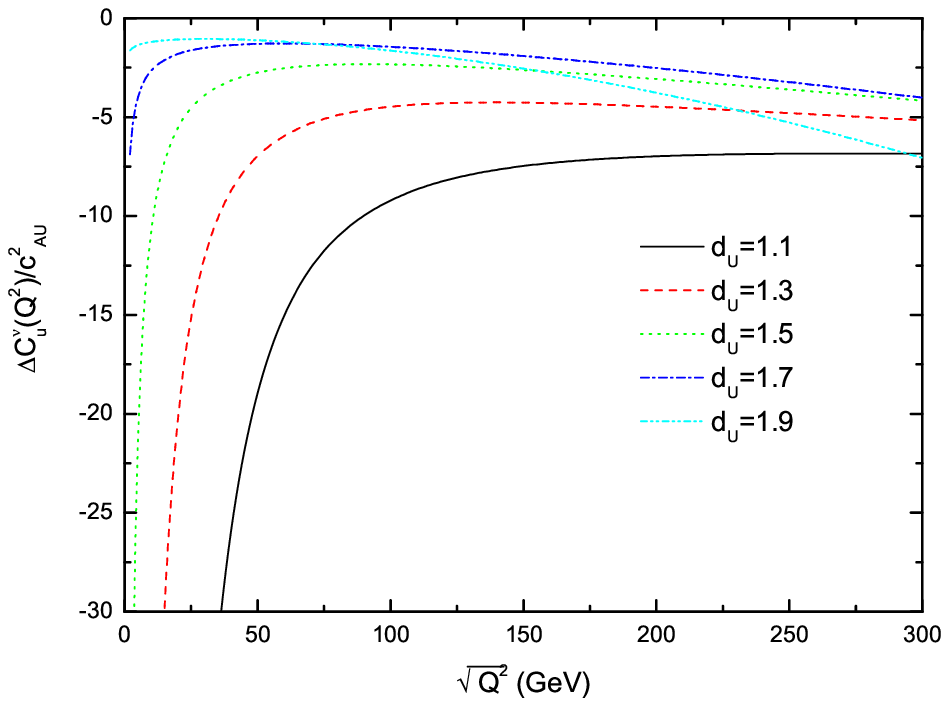}
\caption{The unparticle correction to $C^{\nu}_u(Q^2)$ in unit of
$c^2_{A{\cal U}}$ for $d_{\cal U}=1.1, 1.3, 1.5, 1.7, 1.9$ in the
$c_{V{\cal U}}=0$ and $c_{A{\cal U}}\neq0$ case}
\end{center}
\label{fig4}
\end{figure}

\begin{figure}[hptb]
\begin{center}
\includegraphics*[width=8cm]{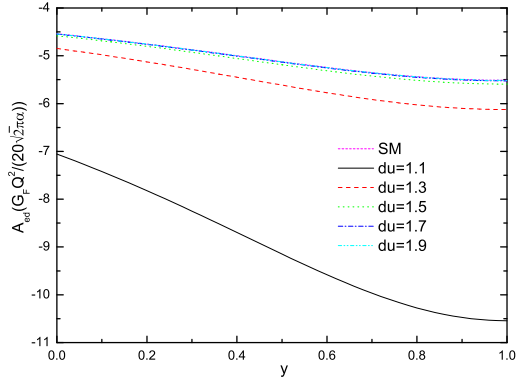}
\caption{The asymmetry
$A_{ed}=\frac{\sigma_R-\sigma_L}{\sigma_R+\sigma_L}$ for the deep
inelastic polarized electron deuteron scattering as a function of
$y=(E-E')/E$ in SM and the new physics scenario with unparticle for
$d_{\cal U}=1.1, 1.3, 1.5, 1.7, 1.9$. The asymmetry is given in unit
of $\frac{G_FQ^2}{20\sqrt{2}\pi\alpha}$, here we assume
$Q^2=10\rm{GeV}^2,c_{V{\cal U}}=c_{A{\cal U}}=0.01$.}
\end{center}
\label{fig5}
\end{figure}

\begin{figure}[hptb]
\begin{center}
\includegraphics*[width=8cm]{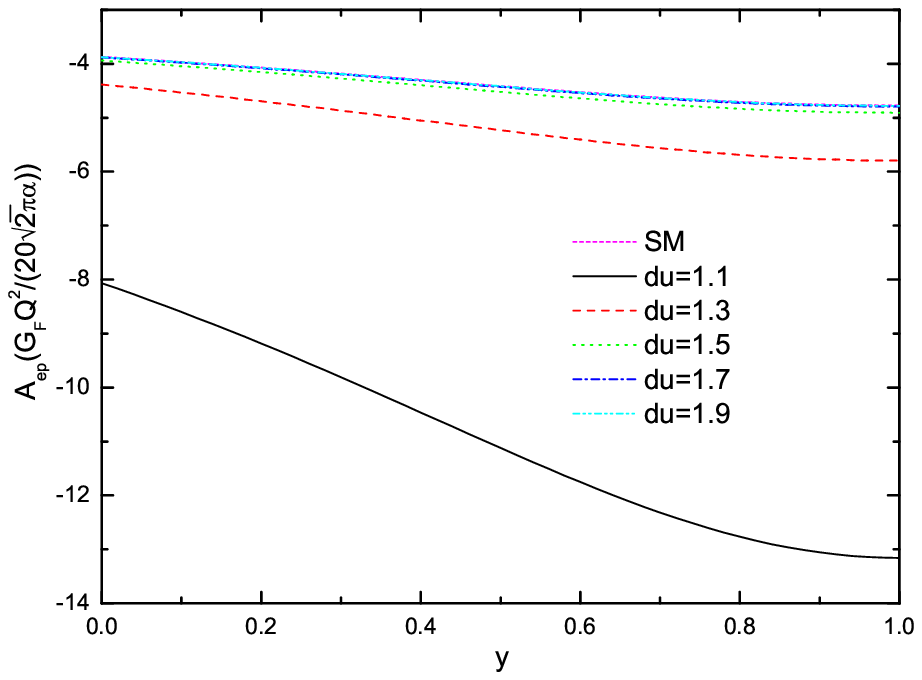}
\caption{The asymmetry
$A_{ep}(x=\frac{1}{3},y)=\frac{\sigma_R-\sigma_L}{\sigma_R+\sigma_L}$
for the deep inelastic polarized electron-proton scattering as a
function of $y=(E-E')/E$ in SM and the new physics scenario with
unparticle for $d_{\cal U}=1.1, 1.3, 1.5, 1.7, 1.9$. The asymmetry
is given in unit of $\frac{G_FQ^2}{20\sqrt{2}\pi\alpha}$, and we
assume $Q^2=10\rm{GeV}^2,c_{V{\cal U}}=c_{A{\cal U}}=0.01$ as in
Fig.5 for the electron-deuteron case.}
\end{center}
\label{fig6}
\end{figure}


\end{document}